# Simulation of RF Cavity Dark Current in Presence of Helical Magnetic Field

GennadyRomanov, Vladimir Kashikhin

## Abstract.

   In order to produce muon beam of high enough quality to be used for a Muon Collider, its large phase space must be cooled several orders of magnitude. This task can be accomplished by ionization cooling. Ionization cooling consists of passing a high-emittance muon beam alternately through regions of low Z material, such as liquid hydrogen, and very high accelerating RF cavities within a multi-Tesla solenoidal focusing channel. But first high power tests of RF cavity with beryllium windows in solenoidal magnetic field showed a dramatic drop in accelerating gradient due to RF breakdowns. It has been concluded that external magnetic fields parallel to RF electric field significantly modifies the performance of RF cavities. However, magnetic field in Helical Cooling Channel has a strong dipole component in addition to solenoidal one. The dipole component essentially changes electron motion in a cavity compare to pure solenoidal case, making dark current less focused at field emission sites. The simulation of dark current dynamic in HCC performed with CST Studio Suit is presented in this paper.

## Introduction.

   The concept of a Muon Collider has been under study internationally for a number of years. In order to produce muon beam of high enough quality to be used for a collider, its large phase space must be cooled several orders of magnitude and done quickly, because of the short life time of the muon. This task can be accomplished by ionization cooling. Ionization cooling consists of passing a high-emittance muon beam alternately through regions of low Z material, such as liquid hydrogen, and very high accelerating RF cavities within a multi-Tesla solenoidal focusing channel. As the particles pass through the low Z material, they lose momentum in 3-D phase space. The longitudinal component is then restored by the RF cavities. There are several designs of ionization cooling channels, and Helical Cooling Channel is one of them. Important to any demonstration of muon ionization cooling is the accelerating cavity technology. Some specific muon beam parameters make closed-cell RF cavity more effective to use in cooling channels. But first high power tests of closed RF cavity with beryllium windows in solenoidal magnetic field showed a dramatic drop in accelerating gradient due to RF breakdowns [1]. It has been concluded that external magnetic fields parallel to RF electric field significantly modifies the performance of RF cavities by deflecting and focusing electrons coming off the surface at field emission sites or shaping any plasma that might form near the surface. That is a serious problem for cooling schemes similar to MICE in which RF cavities are inside big solenoids. However, magnetic field in Helical Cooling Channel has a strong dipole component in addition to solenoidal one. The dipole component essentially changes electron motion in a cavity compare to pure solenoidal case, making dark current less focused at field emission sites. The simulation of dark current dynamic in HCC performed with CST Studio Suit is presented in this paper.

## Model for electromagnetic field simulation.

   The Helical Cooling Chanel design described in [2] has been used as a base for CST model (see Fig.1). The design is very convenient for modeling because it has separate solenoidal and dipole magnets, so it allows independent simulating and manipulating either magnetic field.
   The HCC parameters used in this paper (fields have been slightly corrected compare to the original design) are:

|  |  |
|---|---|
| Period of helix | – 1 m |
| Equilibrium orbit radius | - 0.16 m |
| Solenoidal field | - 5.76 T |
| Helical dipole field | - 1.57 T |

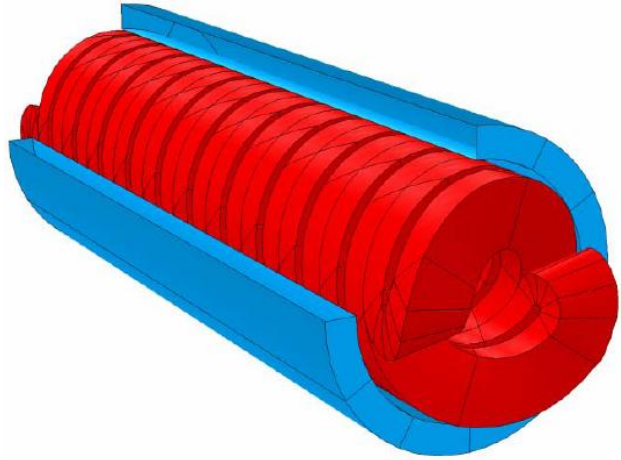

Figure 1. High-field helical dipole with straight solenoid (a quarter removed for clarity)

CST model of the helical dipole consists of two spiral conductors with currents flowing in opposite directions as shown in Fig.2. The transverse cross-section of the conductors represent 120° arc of cylinder. Fig.2 also shows dipole magnetic field simulated with this model in transverse plane.

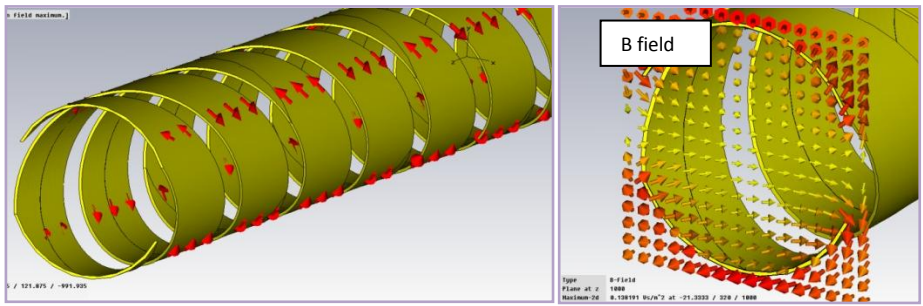

Figure.2. CST model of helical dipole. Arrows show the direction of currents.

CST has a convenient option of using analytical expressions to add fields of simple structure into a model. So, the solenoidal field has been introduced in this way, and there was no need to simulate solenoidal magnet.

To verify the model the muon tracking have been performed in the combined fields of design magnitude. Fig.3 shows the trajectories of muons with kinetic energy of 250 MeV that move along stable helical trajectories with design period, pitch and radius.

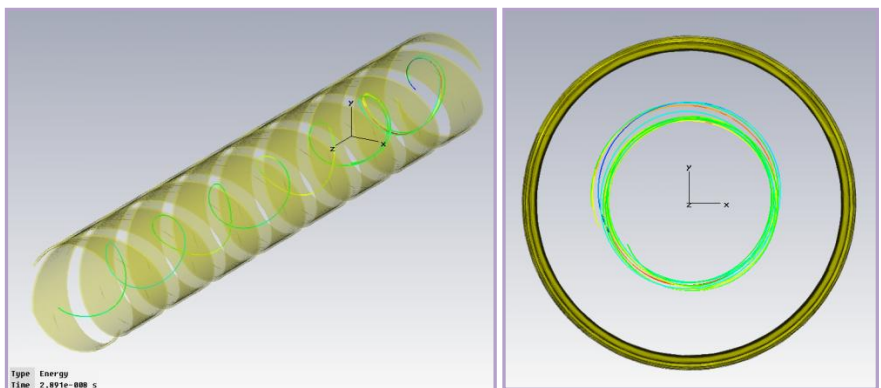

Figure.3. Muon trajectories in the helical cooling channel model (3D and front view).

An 805 MHz RF cavity with beryllium (Be) windows for a muon cooling experiment has been designed, built and tested at high power [3] (see Fig.4). The experiments with this very cavity demonstrated the effect of high solenoidal magnetic field on breakdown gradients. This cavity does not fit mechanically HCC under consideration. But the cavity has inner diameter of 312 mm, and it is good for simulation since we need to use RF volume only. The CST cavity solid model is just slightly offset the helical equilibrium trajectory axis (see Fig.5).

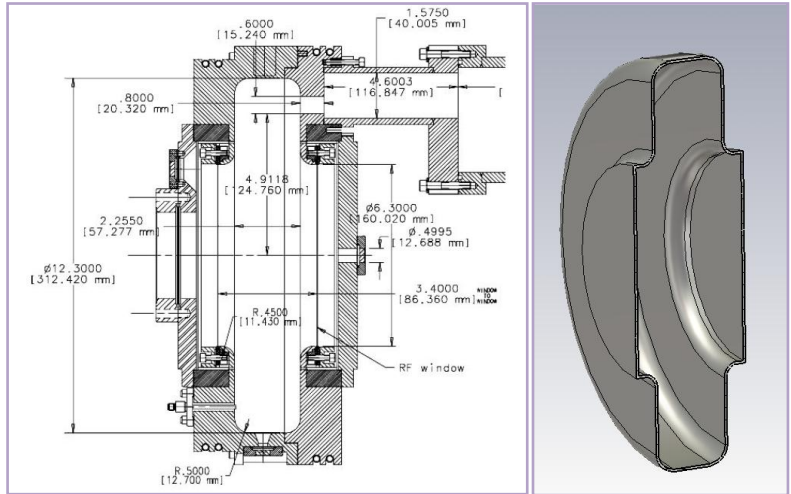

Figure.4. The 805 MHz RF cavity with beryllium (Be) windows design and its CST solid model.

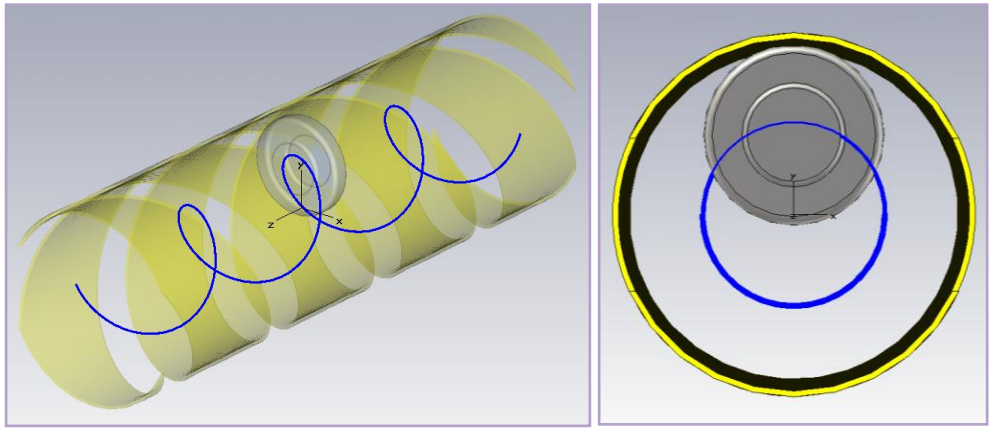

Figure.5. Pill box 805 MHz RF cavity in the HCC. Design equilibrium orbit is shown.

RF fields in the cavity also have been simulated and accelerating gradient of 30 MV/m has been set for further simulation of electron motion inside the cavity.

### Field emission and re-emission models.

In all known models of RF breakdown it is assumed that breakdown in accelerator structure is initiated at field emissions sites in regions of high RF electric field. High field emitted electrons or dark current electrons are produced by tunneling through the electrostatic potential at the surface of the metal, first described in [4]. The sequence of events after significant dark current is developed may be different in different RF breakdown models, but it is clear that dark current dynamic is of primary importance.

To start dark current simulation we have to find the regions with high RF electric field first. The real 805 MHz cavity has two spots of high electric surface field – iris (a traditional spot for accelerating structures) and a sharp edge of coupling slot (see Fig.6). In this study we ignore the spot on the coupling slot since it is a result of given particular design. It definitely needs to be investigated, but in this study we remove the slot out of the model and focus on more general high field region at the cavity iris.

Next step is to create a particle source with certain emission properties in the high field region. CST Studio Suite has well developed model of high field emission particle source, but this option is not convenient for the study by some reasons. And actually there is no necessity to use it. Instead we assigned particle source with fixed emission properties to the iris area. This particle source produces a burst of electrons at the moment of highest electric field. The emitted electrons are uniformly distributed over the iris, with initial energy uniformly distributed from 0 to 4 eV, and uniform angular spread within normal and 45° (see Fig.7).

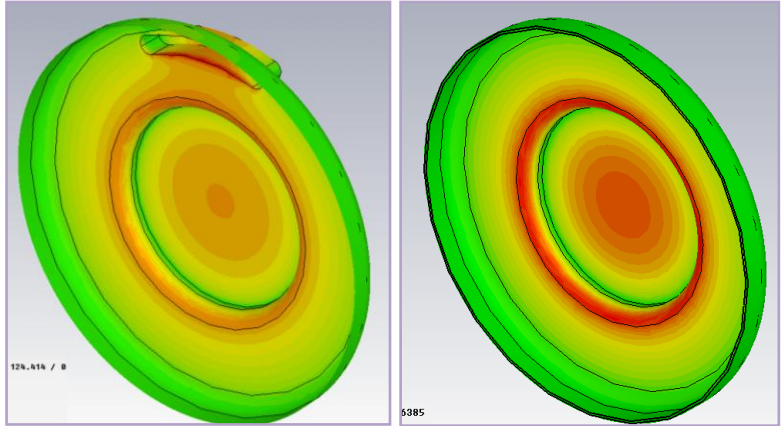

Figure 6. Distribution of RF electric surface field amplitude in the cavity with (left) and without coupling slot.

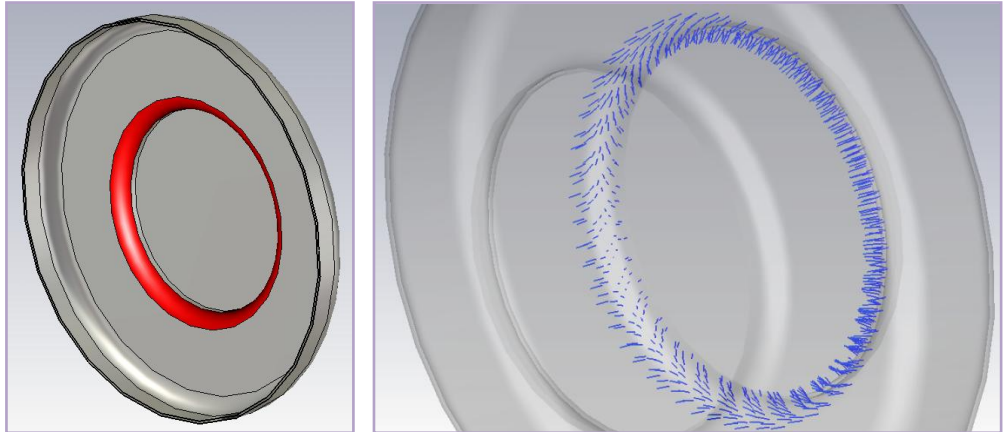

Figure 7. Particle source on the cavity iris (left). Electron trajectories just after emission (only RF fields are on).

Eventually the dark current electrons would hit the cavity walls, and generally they can produce secondary electrons. To gain better insight into the dark current dynamics inside the cavity, the re-emission properties have been assigned to the cavity metal walls. The re-emission process allows us to include secondary emitted electrons into the simulation in addition to the primary dark current electrons

CST Studio Suite uses an advanced secondary emission model based on a probabilistic, mathematically self-consistent theory developed by Furman and Pivi [5]. In this model there are three basic types of secondary electrons: the elastically reflected, the re-diffused and the true secondary ones. A primary electron with energy $E_p$ incident upon an interface will either be scattered elastically without penetration into material (departure energy $E_d = E_p$) or be scattered from one or more atoms inside the material and be reflected (re-diffused) back out ($E_d = 0.2 \div 0.99\ E_p$), or be transmitted across the interface. The transmitted electrons can then excite several electrons within the material. These true secondary electrons travel diffusively and, if close enough to the surface, escape ($E_d \approx$ 2-4 eV).

The model was assumed to be made of copper entirely and corresponding re-emission parameters have been assigned to the walls. The real cavity has TiN coated beryllium windows, so the emission property is reduced for these surfaces. But the windows are not involved significantly in dark current generation, and the copper windows should not change much the dark current dynamic.

**Dark current simulation in solenoidal field.**

Simulation of dark current in RF cavity in presence of solenoidal field only is useful by three reasons: 1) this is an important real case; 2) the dark electron trajectories can be easily checked analytically and therefore the model can be verified; 3) the case is a good reference point for understanding of dark current particle dynamic.

First let us have a look at dark current dynamic with RF filed on only and without any static magnetic field that is shown in Fig.8.

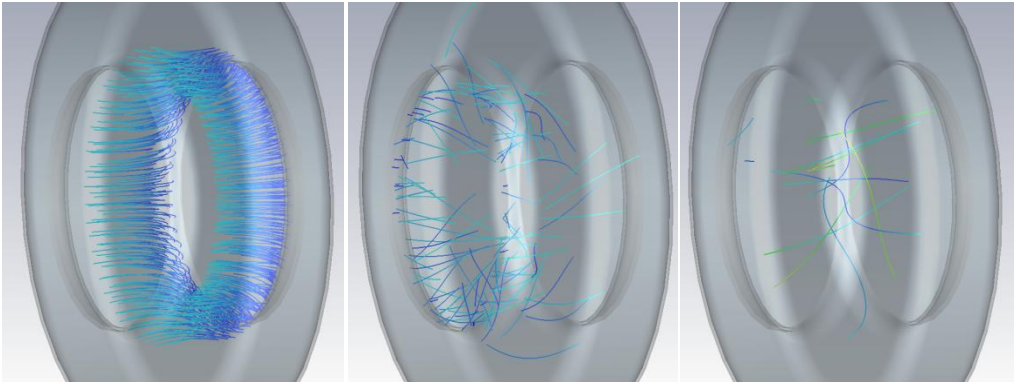

Figure.8. Snapshots of dark current electron trajectories at 0.23 ns, 0.5 ns and 1.3 ns after emission (the trajectories of "survived" electrons only are plotted)

Initial number of emitted electrons of 600 is intentionally relatively small to show more clearly the character of electron motion. During first RF period, just after initial emission, the electrons are moving in more or less compact group. When they hit opposite iris and wall some of them are re-diffused and elastically back scattered. The electrons are mostly backscattered because the probability for generation of true secondaries is very low, since the average energy of electrons at the first hit is rather high ≈1.1 MeV. The reflected electrons acquire angular (±45°) spread and significant energy spread from 0 to maximum of incident energy. So, the electrons of second generation are already very much dispersed and motion of the next generations is completely chaotic. It is important to notice that some of the elastically scattered electrons are accelerated up to 4.4 MeV, and therefore they are more capable to damage the copper surface.

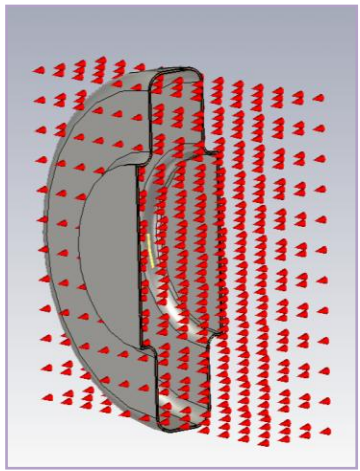

Figure 9. RF cavity in solenoidal field.

To simulate dark current in RF cavity placed inside magnets, the whole magnetic field map is not needed. Instead only cut of map that fit RF cavity is used (see Fig.9). The simulations show that starting with solenoidal field of 1 T, the dark current electrons move strictly along magnetic lines of force and hit practically mirror image spot on the opposite iris. Moreover the backscattered electrons also move back along solenoidal field due to very strong focusing in spite of angular and energy spread and hit the original emission site (see Fig.10). Number of backscattered electron drops very quickly during 3-4 RF periods. But initial high field emitted electrons are originated at each RF period, so total dark current may even increase with time. Average impact energy is 0.4 MeV, which is lower than that without magnetic field, but it is still too high for true secondaries generation. Maximal impact energy is ≈ 3 MeV for the elastically scattered electrons. At these energies the relative velocity of backscattered electrons reaches β=0.8÷0.99, which is close to the design β for muons. It means that major portion of backscattered electrons move almost in sync with RF electric field, gaining higher energy and increasing the danger of the dark current impact.

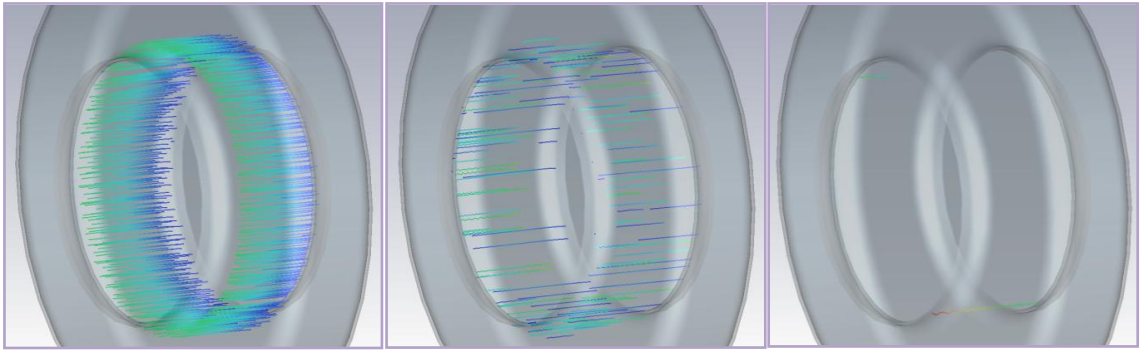

Figure 10. Snapshots of dark current electron trajectories at 0.23 ns, 0.5 ns and 1.3 ns after emission in presence of 5.7 T solenoidal field.

The principal difference between the two dark current dynamics is that the impacts are spread over larger area in the absence of solenoidal field, while with magnetic field the impacts are focused to small spots, which are actually initial field emission sites. The difference is illustrated in Fig. 11, where the impacts are shown in XY phase space in the location of iris for both cases.

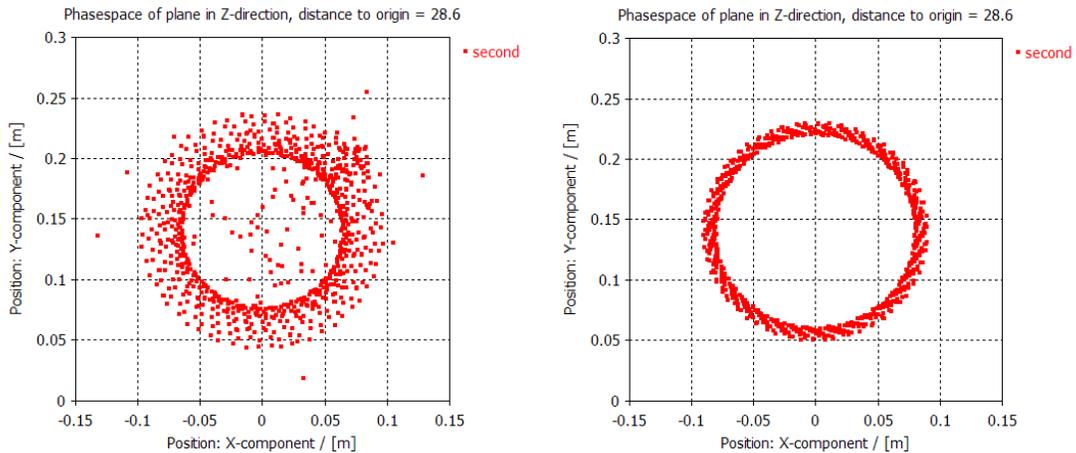

Figure 11. Impact coordinates in transverse phasespace with (right) and without (left) solenoidal magnetic field.

A number of models have been proposed for RF breakdown with and without a magnetic field present. All models consider high field emission intense bombardment near a field emission site with high surface field as a most probable trigger for breakdown or at least as an essential component of breakdown mechanism. Therefore a dark current being well focused and synchronized with RF fields explains gradient limitations of pillbox cavity in strong solenoidal magnetic field.

Strongly focused dark current beam with straight line electron trajectories seems to be inconsistent with observed Cu droplets spreaded over broad area. One may assume that some mechanism, such a space charge, spreads the beamlets out to a greater extent than the electron beam image. Indeed, intense field emitted electron beam damages the surface and creates new asperities with higher field enhancement coefficient. As a result the dark current increases significantly and some space charge effects may be expected. But our simulation of electron motion with dark current up to 100 A did not show noticeable effect – the electron beam remained well focused. On the other hand the products of breakdown events are very different and their trajectories are complex. While neutral atoms, fragments and clusters fractured or evaporated off the metal surface will all fly ballistically into the cavity, the fate of ions injected into the cavity volume is somewhat more complex. The ions will be affected by the high EM fields, by field emitted electron beams and by the initial parameters of the emission site that produced them. Large droplets with low charge to mass ratio would move slowly and stay in range of the emitter [6]. Besides, as it was mentioned above, field distribution in the real cavity has azimuthal asymmetry which can make Cu droplets deposition very nonuniform. This is consistent with the examination of the inside of the tested cavity (Fig.12) that showed severe pitting on the irises and an asymmetric spray of Cu over the windows surface and Cu powder in the bottom [1].

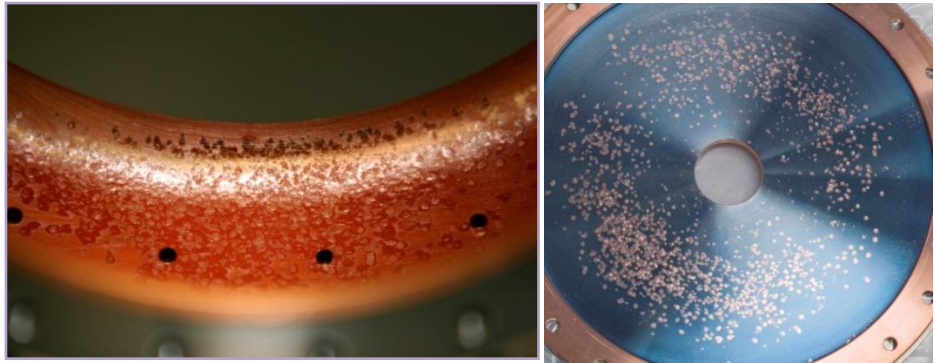

Figure 12. Damaged iris surface and Cu deposition on the Be window.

We skipped simulation of dark current emitted from the edge of coupler slot, where surface electric field is maximal and enhanced by the rather sharp edge. But the mechanism of breakdown is quite similar to that at irises. The electrons emitted from the edge fly along straight lines being focused by magnetic field and strike the opposite wall, producing damaged area resembling the edge shape.

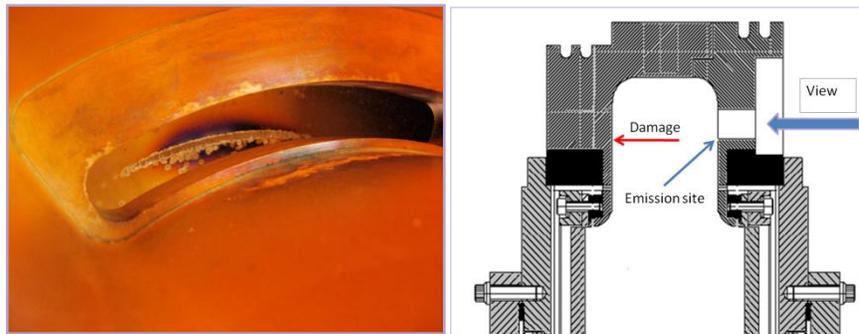

Figure 13. Surface damage due to the dark current emission from the coupling slot.

### Dark current simulation in combined solenoidal and dipole fields of HCC.

For dark current simulation in combined solenoidal and dipole fields we used again only parts of magnetic field maps. The cut of dipole field map from the helical dipole model (the model is shown in Fig.2) has been added to RF and solenoidal magnetic fields (See Fig.14).

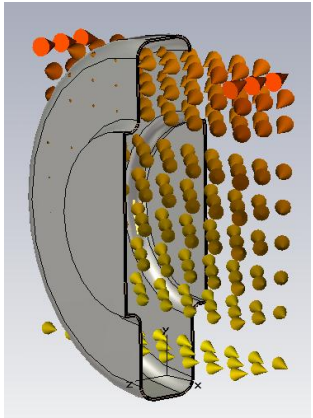

Figure 14. The helical dipole field cut used in simulations.

Particle source that simulates initial high field emitted electrons from the iris is the same and is shown in Fig.7. After emission the electrons again move along magnetic force lines, but now the lines are bent in spiral, so the trajectories look as it is shown in Fig.15.

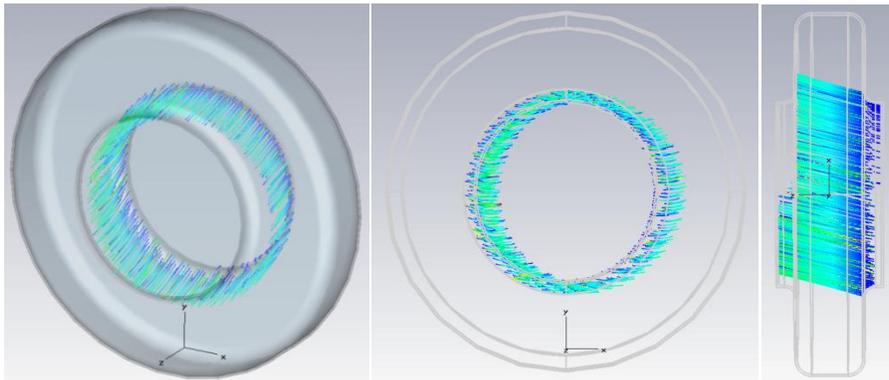

Fig.15. Snapshots of dark current electron trajectories at 0.23 ns after emission in presence of 5.7 T solenoidal field and helical dipole field of 1.57 T on HCC axis.

The impact coordinates in the transverse phase space show that dark current is still well focused, but misses opposite iris almost completely (Fig.16).

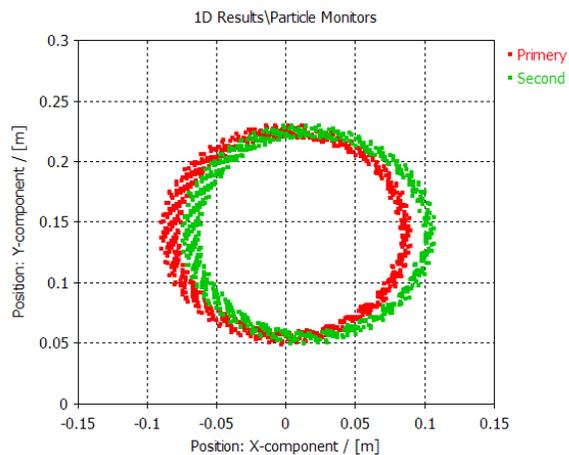

Figure.14. Impact coordinates in transverse phasespace of the cavity in HCC fields.

This is an important change in dark current dynamic since now there is no high field emission intense bombardment near a field emission site with high RF electric surface field. The dark current electron energy reaches 1-3 Mev, when they hit the walls, and the electrons still can damage the surface, creating craters and asperities. But the locations of such damages are at low gradient areas, so there is neither immediate breakdown nor dark current increase.

One more thing may reduce dark current generation. Approximately 50% of dark current electrons generated at the iris hit the beryllium window. The beryllium windows have TiN coating and therefore reduced true secondary emission yield. The coating eliminates very effectively low energy multipacting that is based on true electron re-emission. If the TiN coating also reduces the elastic scattering, which dominates in our case, then it could be an additional dumping of dark current generation. Unfortunately no data on elastic scattering of TiN has been found to make more definite conclusion.

To estimate probable decline of secondary emission while more and more electrons hit beryllium windows instead of copper, it was assumed that TiN coated beryllium does not have elastic scattering at all. At fixed nominal magnetic solenoidal and RF fields the strength of magnetic dipole field was scanned and dark current simulation was performed to track changes in integral secondary emission. The simulation showed 40% decline of secondary emission at the moment when approximately half of the primary electrons miss copper walls. It means that smaller number of secondary electrons returns back and hits field emission site.

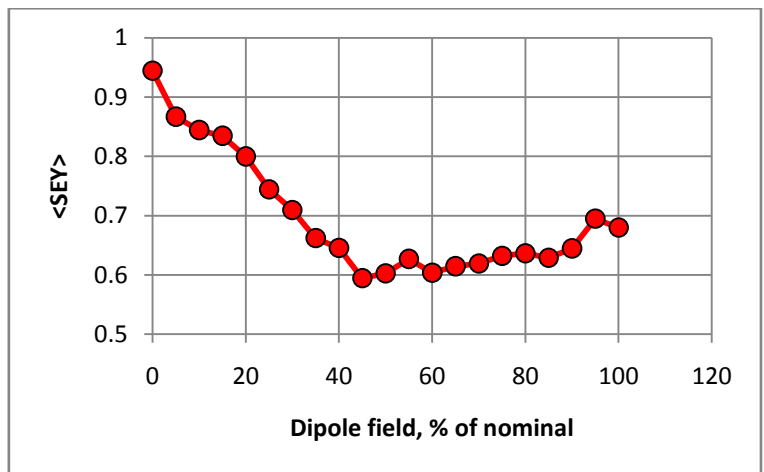

Figure 15. Integral secondary emission vs dipole field strength.

## Conclusion

The dark current in the RF accelerating cavity in Helical Cooling Channel has been simulated. The simulation of electron motion in combined RF, magnetic solenoidal and magnetic dipole fields revealed qualitative change in the dark current dynamic compare to pure solenoidal magnetic field. That is the dark current is still well focused, but misses opposite iris almost completely. Therefore there is no high field emission intense bombardment near a field emission site with high RF electric surface field. Along with probable reduction of total number of elastically scattered electrons that can diminish the problem of RF breakdowns in accelerating cavities in HCC.